\documentclass{revtex4}

\usepackage{graphicx}
\usepackage{bm}
\usepackage{amssymb}
\usepackage{amsmath}

\begin{document}

\title{Quantum support to BoHua Sun's conjecture}

\author{Claude \surname{Semay}}
\email[E-mail: ]{claude.semay@umons.ac.be}
\affiliation{Service de Physique Nucl\'{e}aire et Subnucl\'{e}aire,
Universit\'{e} de Mons,
UMONS Research Institute for Complex Systems,
Place du Parc 20, 7000 Mons, Belgium}
\date{\today}

\begin{abstract}
A generalization of the Kepler's third law has been proposed by BoHua Sun for $N$-body periodic orbits in a Newtonian gravitation field. In this paper, it is shown that this formula can apply for a quantum system of $N$ self-gravitating identical particles, for a good choice of the period for the quantum motion. 
\end{abstract}

\maketitle

The Kepler's third law is certainly a milestone in the development of mechanics. But this peculiar relation between the period and the size of the orbit (or the period and the energy of the orbit) seemed only relevant for two-body systems. Nevertheless, the computation of a great number of Newtonian periodic planar collisionless three-body orbits suggests the existence of a quasi Kepler's third law \cite{li17}. A definition of an average period 
\begin{equation}
\label{Tb}
\bar T=\frac{T}{L_f}
\end{equation}
must be chosen, with $T$ the period of the orbit and $L_f$ the length of the so-called ``free group element" of the orbit. This corresponds to a method to classify the topology of the orbit around the three two-body collision points \cite{suv14}. The scale-invariant average period 
\begin{equation}
\label{Ts}
T_*=\bar T\, |E|^{3/2},
\end{equation}
where $E$ is the total energy (kinetic and potential) of the orbit is found by  numerical computation approximately equal to a universal constant for three identical bodies \cite{li17}. Using arguments based on dimensional analysis, BoHua Sun suggested then a generalization of the Kepler's third law for $N$-body periodic orbits \cite{sun18}:
\begin{equation}
\label{BHS}
T_*=\frac{\pi}{\sqrt{2}}\,G\, \left( \frac{\sum_{i=1}^N \sum_{j=i+1}^N (m_i m_j)^3}{\sum_{i=1}^N m_i}\right)^{1/2}.
\end{equation}
This formula reduces to the usual Kepler's third law for $N=2$, and gives results compatible with computations based on a great number of  periodic planar collisionless orbits for $N=3$ \cite{li17,li17b,li18}. More precisely, the agreement is good when one of the mass is much larger than the two others, and poor when one of the mass is much smaller than the two others. In particular, for identical bodies, formula~(\ref{BHS}) reduces to
\begin{equation}
\label{BHSid}
T_*=\frac{\pi}{2}\,G\,m^{5/2}\,\sqrt{N-1}.
\end{equation}

Although macroscopic and microscopic worlds are very different, strong connections exist between classical and quantum theories. The most famous one is certainly the Ehrenfest theorem, showing that expectation values obey Newton's second law. Moreover, three fundamental theorems of quantum mechanics (Hellmann-Feynman theorem, virial theorem and comparison theorem) have classical counterparts \cite{sema18a}. So, a natural question arises: Is the Kepler's third law can also be relevant for quantum $N$-body systems? Such systems are not easy to treat, but the problem is less difficult for identical particles. In this case, the necessity to (anti)symmetrize the wave-function puts strong constraints on the solution, which can help the computation. The envelope theory is an efficient and convenient tool to obtain approximate solutions for quantum systems made of $N$ identical particles \cite{hall80,sema13a,sema18b}. With this method, the quantum energy of $N$ self-gravitating particles with a mass $m$ in $D$-dimensional space is given by \cite{sema15a}   
\begin{equation}
\label{E}
E_q= - \frac{N^2(N-1)^3}{16} \frac{G^2\, m^5}{Q^2\, \hbar^2},
\end{equation}
where 
\begin{equation}
\label{QN}
Q = \sum_{i=1}^{N-1} (2 n_i + l_i) + (N - 1)\frac{D}{2}
\end{equation}
is a global quantum number (corresponding to $N-1$ identical harmonic oscillators). The equivalent of the quantity $\bar T$ for this quantum system is not easy to define, since the dynamics is described by a wave-function in the configuration space and not by motions of particles through the real space. Nevertheless, the equations of the envelope theory can be interpreted within a semi-classical framework \cite{sema13a,sema18b} in the following way: the $N$ particles are in circular motion with the same
momentum $p_0=Q\, \hbar/r_0$ (and thus the same speed $v_0=p_0/m$) at a distance $d_0=r_0/N$ from the center of mass, each particle
being at an angular distance of $2\,\pi/N$ from its neighbors. The distance $r_0$ for self-gravitating particles is given by \cite{sema15a}
\begin{equation}
\label{r0}
r_0= \frac{2^{3/2}}{N^{1/2}(N-1)^{3/2}} \frac{Q^2\, \hbar^2}{G\, m^3}.
\end{equation}
Within this picture, the quantum period $T_q$ can be estimated by the time taken by a particle to run a complete turn 
\begin{equation}
\label{Tq}
T_q=\frac{2\,\pi\,d_0}{v_0}=2\,\pi \frac{r_0}{N} \frac{m}{p_0}.
\end{equation}
Formulas (\ref{E}) and (\ref{Tq}) give
\begin{equation}
\label{Tq}
T_q\, |E_q|^{3/2} = \frac{\pi}{4}\,G\,m^{5/2}\,N\,(N-1)^{3/2}
\end{equation}
This quantity is $N(N-1)/2$ times $T_*$ given by (\ref{BHSid}). $N(N-1)/2$ is the number of pairs but also the number of two-body collision points in the $N$-body systems. So, if we assume that the period must be rescaled by this number, the quantum scale-invariant average period is given by 
\begin{equation}
\label{Tsq}
T_{*q}=\frac{2\, T_q}{N(N-1)}\, |E_q|^{3/2}
\end{equation}
for $N$ identical particles. Then, $T_{*q}$ and $T_*$ give both the same result. 

One can ask how is it possible that an approximate quantum formula gives the same result than a classical (assumed) exact relation? First, the quantum character of formulas (\ref{E}) and (\ref{r0}) is carried by the quantity $Q\,\hbar$, which disappears in the computation of $T_{*q}$. Second, the main source of error in the calculations from the envelope theory is the strong degeneracy (inherent to the method) due to the quantum number $Q$. The results can be greatly improved by modifying the structure of $Q$, either by a fit on numerical accurate computations \cite{sema15a} or by the use of a complementary method \cite{sema15b}. As mentioned above, this quantum number disappears in the computation of $T_{*q}$. 

It is interesting to look at the quantum two-body Coulomb system for which exact results are well-known. If the Hamiltonian is written
\begin{equation}
\label{Hcoul2}
H=\frac{\bm p^2}{2\mu}-\frac{\kappa}{r},
\end{equation}
and if we take $d_0=\langle 1/r \rangle^{-1}$ and $v_0=\sqrt{\langle \bm p^2 \rangle}/\mu$, then
\begin{equation}
\label{tscoul2}
T_{*q} = \frac{\pi}{\sqrt{2}}\,\mu^{1/2}\,\kappa,
\end{equation}
with obviously only one two-body collision point. If $\mu=m_1\,m_2/(m_1+m_2)$ and $\kappa=G\,m_1\,m_2$, (\ref{BHS}) is recovered with $N=2$. With a good and reasonable choice for the period, classical and quantum results are again identical.

Let us mention that a linear dependence seems to exist between $T\, |E|^{3/2}$ and a number representing the orbit's topological complexity for collisionless periodic three-body orbits in the Coulomb potential with one positively charged particle and two negatively charged particles \cite{sind18}. The existence of attractive and repulsive interactions in this system makes it different from the ones presented above. Nevertheless, the dependence found could be an indication of some universal relation for Coulombic systems.

I suspect that the identical classical and quantum results obtained for a generalization of the Kepler's third law for $N$-body systems is more than a simple happy coincidence, even if the ``quantum period" must be well chosen. But this problem certainly deserves more studies. A possible way is to search for regularities in Newtonian mechanics for more than three orbiting bodies. Another way is to generalize the envelope theory for different particles and look for new invariants.


\begin{thebibliography}{99} 

\bibitem{li17} X.M. Li and S.J. Liao, More than six hundred new families of Newtonian periodic planar collisionless three-body orbits, Sci. China-Phys. Mech. Astron. \textbf{60}, 129511 (2017)
\bibitem{suv14} M. \v{S}uvakov and V. Dmitra\v{s}inovi\'{c}, A guide to hunting periodic three-body orbits, Am. J. Phys. \textbf{82}, 609 (2014)
\bibitem{sun18} B.H. Sun, Kepler's third law of $n$-body periodic orbits in a Newtonian gravitation field, Sci. China-Phys. Mech. Astron. \textbf{61}, 054721 (2018)
\bibitem{li17b} X.M. Li, Y.P. Jing, and S.J. Liao, The 1223 new periodic orbits of planar three-body problem with unequal mass and zero angular momentum, arXiv:1709.04775.
\bibitem{li18} X.M. Li and S.J. Liao, Collisionless periodic orbits in the free-fall three-body problem, arXiv:1805.07980
\bibitem{sema18a} C. Semay, Three theorems of quantum mechanics and their classical counterparts, Eur. J. Phys. \textbf{39}, 055401 (2018)
\bibitem{hall80} R.L. Hall, Energy trajectories for the $N$-boson problem by the method of potential envelopes, Phys. Rev. D \textbf{22}, 2062 (1980)
\bibitem{sema13a} C. Semay and C. Roland, Approximate solutions for $N$-body Hamiltonians with identical particles in $D$ dimensions, Res. in Phys. \textbf{3}, 231 (2013)
\bibitem{sema18b} C. Semay and G. Sicorello, Many-body forces with the envelope theory, Few-Body Syst. \textbf{59}, 119 (2018)
\bibitem{sema15a} C. Semay, Numerical Tests of the Envelope Theory for Few-Boson Systems, Few-Body Syst. \textbf{56}, 149 (2015)
\bibitem{sema15b} C. Semay, Improvement of the envelope theory with the dominantly orbital state method, Eur. Phys. J. Plus \textbf{130}, 156 (2015)
\bibitem{sind18} M. \v{S}indik, A. Sugita, M. \v{S}uvakov  and V. Dmitra\v{s}inovi\'{c}, Periodic three-body orbits in the Coulomb potential, Phys. Rev. E \textbf{98}, 060101(R) (2018)

\end{thebibliography}
\end{document}